\documentclass[11pt]{report}
\usepackage{hyperref}
\usepackage{amsmath}
\usepackage{cleveref}
\newcommand{\HRule}{\rule{\linewidth}{0.5mm}}
\renewcommand*{\thefootnote}{\fnsymbol{footnote}}

\begin{document}
\hypersetup{pageanchor=false}

\thispagestyle{empty}

	\begin{flushright}
	  {\large
            \textbf{\href{https://https://cds.cern.ch/record/2886099}{LHCHWG-2024-001}}} \\[0.5cm]	
		{\large 	\textrm{May 2, 2023}} \\[2.0cm]
	\end{flushright}

	\begin{center}

	\textsc{\Large 	\href{https://twiki.cern.ch/twiki/bin/view/LHCPhysics/LHCHWG}{LHC Higgs Working Group}\footnote{\href{https://twiki.cern.ch/twiki/bin/view/LHCPhysics/LHCHWG}{\sl https://twiki.cern.ch/twiki/bin/view/LHCPhysics/LHCHWG}}} \\[0.5cm]
	\textsc{\Large 	Public Note} \\[1.5cm]
	
	\HRule \\[0.9cm]
	\textbf{\Large Ad interim recommendations for the Higgs boson production cross sections at $\sqrt{s}=13.6$ TeV} \\[1.0cm]
	\HRule \\[1.5cm]

	\textrm{
    {\large {\bf WG1 conveners:} Alexander Karlberg$^{1}$,
     Julie Malcles$^{2}$,
     Bernhard Mistlberger$^{3}$,
     Roberto~Di~Nardo$^{4}$ \\
     {\bf $\boldsymbol{ggF}$ conveners:}  Syed Haider Abidi$^{5}$, Robin Hayes$^{6}$, Alexander Huss$^1$, Stephen Jones$^{7}$,\\ 
     {\bf VBF conveners:} Gaetano Barone$^{8}$, Jiayi Chen$^{9}$, Stephane~Cooperstein$^{10}$, Silvia Ferrario Ravasio$^{1}$, Mathieu Pellen$^{11}$ \\
     {\bf $\boldsymbol{VH}$ conveners:} Hannah Arnold$^{6}$, Alessandro Calandri$^{12}$, Suman Chatterjee$^{13}$, Giancarlo~Ferrera$^{14}$, Ciaran Williams$^{15}$\\ 
     {\bf $\boldsymbol{t\bar{t}H/tH}$ conveners:} Malgorzata Worek$^{16}$, Marco Zaro$^{14}$\\ 
     {\bf $\boldsymbol{b\bar{b}H/bH}$ conveners:} Chayanit Asawatangtrakuldee$^{17}$, Tim Barklow$^{3}$, Michael Spira$^{18}$, Marius~Wiesemann$^{19}$
    }} \\[0.3cm]	
    \textit{$^{1}$CERN, Theoretical Physics Department, CH-1211 Geneva 23, Switzerland } \\
    \textit{$^{2}$IRFU, CEA, Universit\'e Paris-Saclay, Gif-sur-Yvette, F-91191, France} \\
    \textit{$^{3}$SLAC National Accelerator Laboratory, Stanford University, Stanford, CA 94039, USA } \\
    \textit{$^{4}$Universit\'{a} degli Studi and INFN Roma Tre, Via della Vasca Navale 84, I-00146, Rome, Italy}\\
    \textit{$^{5}$Brookhaven National Laboratory, PO Box 5000, Upton, NY 11973, USA}\\   
    \textit{$^{6}$Nikhef, Science Park 105, 1098 XG Amsterdam, The Netherlands }\\
    \textit{$^{7}$Institute for Particle Physics Phenomenology, Durham University, Durham DH1 3LE, UK}\\
    \textit{$^{8}$Brown University, Providence, RI 02912, USA} \\
    \textit{$^{9}$Department of Physics, Simon Fraser University, Burnaby, BC V5H0G9, Canada}\\
    \textit{$^{10}$University of California San Diego, 500 Gilman Dr, La Jolla, CA 92093, USA}\\
    \textit{$^{11}$Physikalisches Institut, Albert-Ludwigs-Universit\"at Freiburg, D-79104 Freiburg, Germany} \\
    \textit{$^{12}$ETH Zurich—Institute for Particle Physics and Astrophysics (IPA), Zurich, Switzerland}\\
    \textit{$^{13}$Institute of High Energy Physics (HEPHY), Austrian Academy of Sciences (\"{O}AW), Nikolsdorfer Gasse 18, 1050 Vienna, Austria}\\
    \textit{$^{14}$Universit\'{a} degli Studi di Milano $\&$ INFN, Sezione di Milano, Via Celoria 16, 20133 Milano, Italy}\\
    \textit{$^{15}$Department of Physics, University at Buffalo, The State University of New York, Buffalo, 14260, USA}\\
    \textit{$^{16}$Institute for Theoretical Particle Physics and Cosmology, RWTH Aachen University, \\D-52056 Aachen, Germany}\\
    \textit{$^{17}$Department of Physics, Faculty of Science, Chulalongkorn University, 10330 Bangkok, Thailand}\\
    \textit{$^{18}$Paul Scherrer Institut, CH–5232 Villigen PSI, Switzerland}\\
    \textit{$^{19}$Max-Planck-Institut f\"ur Physik, F\"ohringer Ring 6, 80805 M\"unchen, Germany}\\

\textbf{}
	\end{center}

        \thispagestyle{empty}



\mbox{}\vspace*{3em}
\begin{center}
    \textbf{Abstract}
 This note documents predictions for the inclusive production cross sections of the Standard Model Higgs boson at the Large Hadron Collider at a centre of mass energy of $13.6$ TeV. The predictions here are based on simple extrapolations of previously documented predictions published in the CERN Yellow Report "Deciphering the Nature of the Higgs Sector". The predictions documented in this note should serve as a reference while a more complete and update-to-date derivation of cross section predictions is in progress.
\end{center}

\newpage
\hypersetup{pageanchor=true}
\renewcommand*{\thefootnote}{\arabic{footnote}}
\setcounter{footnote}{0}
\setcounter{page}{1}
\chapter*{Higgs boson production cross sections at $\sqrt{s}=13.6$ TeV}
This documents collects the ad interim recommendation for the Higgs boson inclusive production cross section at $\sqrt{s}=13.6 $ TeV.
These are obtained from extrapolations based on the cross-section values computed at different value of $m_H$ and centre of mass energy in the CERN Yellow Report "Deciphering the Nature of the Higgs Sector" (YR4) (CERN-2017-002)~\cite{LHCHiggsCrossSectionWorkingGroup:2016ypw}.  
The values have been computed within the relevant subgroups of the WG1 LHC Higgs Working Group. 
Consequently, the derived numerical predictions are a direct product of the work of YR4 and the references cited therein, and it is therefore recommended both this note and the YR4 are cited whenever the interim predictions are used in experimental analyses.

\section*{ggF}
The ggF cross sections and uncertainties at different values of $m_H$ are reported in Table ~\ref{tab:ggF} and obtained with a linear interpolation of the YR4~\cite{LHCHiggsCrossSectionWorkingGroup:2016ypw} cross sections available at $\sqrt{s}=13\,\mathrm{TeV}$ and $\sqrt{s}=14\,\mathrm{TeV}$. 
The interpolation formula utilised is,
\begin{equation}
    \sigma(13.6\,\mathrm{TeV})=0.4\times\sigma(13\,\mathrm{TeV})+0.6 \times\sigma(14\,\mathrm{TeV}),
\end{equation}
where $\sigma(E)$ is the production cross section at centre of mass energy $E$. We have checked explicitly by computing a few cross sections exactly, that the error introduced by the linear interpolation is below the permille level.

The YR4 numbers incorporate results from:
\begin{itemize}
    \item QCD corrections in the heavy-top limit at NLO~\cite{Dawson:1990zj,Graudenz:1992pv,Djouadi:1991tka}, NNLO~\cite{Anastasiou:2002yz,Harlander:2002wh,Ravindran:2003um} and N$^3$LO~\cite{Anastasiou:2014lda,Anastasiou:2014vaa,Anastasiou:2015vya}.
    \item Exact finite quark mass effects at NLO~\cite{Graudenz:1992pv,Spira:1995rr}.
    \item Approximate top quark mass effects obtained in a $1/m_t$ expansion at NNLO~\cite{Harlander:2009mq,Pak:2009dg,Harlander:2009bw,Harlander:2009my}.
    \item Electroweak corrections at NLO~\cite{Aglietti:2004nj,Actis:2008ug,Actis:2008ts}
    \item Approximate mixed QCD-electroweak corrections obtained in the $m_t, m_W, m_Z \gg m_H$ limit~\cite{Anastasiou:2008tj}.
\end{itemize}
The combination procedure and error prescription are described in Ref.~\cite{Anastasiou:2016cez}, and numbers are produced using {\sc iHixs}~\cite{Dulat:2018rbf}.

\section*{VBF}

In Ref.~\cite{LHCHiggsCrossSectionWorkingGroup:2016ypw}, the state-of-the-art predictions were obtained at NNLO QCD + NLO EW accuracy.
The NNLO QCD corrections were obtained from {\sc proVBFH}~\cite{Cacciari:2015jma} while the EW corrections have been computed with the help of {\sc Hawk}~\cite{Ciccolini:2007ec,Ciccolini:2007jr,Denner:2014cla}.
Both types of corrections are obtained in the VBF approximation and do not include $s$-channel contributions.
They are combined using the following formula:
\begin{align}
\sigma^{\rm VBF} = \sigma^{\rm DIS}_{\rm NNLO\,QCD} (1+\delta_{\rm EW}) + \sigma_\gamma,
\end{align}
where $\delta_{\rm EW}$ is the relative EW correction and $\sigma_\gamma$ the photon-induced contributions.
The VBF cross sections at $13.6\,{\rm TeV}$ for different values of $m_H$ are reported in Table ~\ref{tab:VBF}.
These are obtained upon applying a cubic spline interpolation based on numbers computed in Ref.~\cite{LHCHiggsCrossSectionWorkingGroup:2016ypw} at various centre of mass energies.

\section*{VH}
The WH and ZH cross sections at different values of $m_H$ are reported in Table ~\ref{tab:WH} and  Table ~\ref{tab:ZH} respectively. These are obtained by first performing a quadratic fit to the available YR4 cross-sections at various centre of mass energies and then extracting the value for $\sqrt{s}=13.6$ TeV. The NNLO-QCD corrections were obtained with {\tt VH@NNLO}~\cite{Brein:2003wg,Altenkamp:2012sx,Brein:2012ne,Harlander:2013mla,Harlander:2014wda} and the NLO-EW with {\sc Hawk}~\cite{Denner:2011id,Denner:2014cla}.

\section*{$\boldsymbol{t\bar{t}H}$ and $\boldsymbol{tH}$}
The $t\bar{t}H$ and $tH$ ($t$-channel, $s$-channel, $tWH$) cross sections at different values of $m_H$ are reported in Table \crefrange{tab:ttH}{tab:tWH} and are obtained with a linear interpolation of the YR4 cross sections. In YR4, cross-sections
for $\boldsymbol{t\bar{t}H}$ and $\boldsymbol{tH}$ have been computed with {\sc MadGraph5\_aMC@NLO}~\cite{Alwall:2014hca,Frederix:2018nkq}, as follows:
\begin{itemize}
\item $t\bar{t}H$ production is computed at NLO QCD+EW accuracy
\cite{Beenakker:2001rj,Reina:2001sf,Beenakker:2002nc,Reina:2001bc,Dawson:2002tg,Frixione:2014qaa,Zhang:2014gcy,Frixione:2015zaa,Frederix:2018nkq}.
\item $t$-channel $tH$ is computed at NLO QCD, as documented in Ref.~\cite{Demartin:2015uha}. In particular, the central value of the cross
section is computed in a five-flavour scheme, but the quoted scale uncertainty
are the envelope of the scale uncertainty obtained both in the five- and four-flavour
scheme; hence, they account also for the flavor-scheme ambiguity.
\item $t$-channel $tH$ is computed at NLO QCD, as documented in Ref.~\cite{Demartin:2015uha}.
\item $tWH$ is computed at NLO QCD, following Ref.~\cite{Demartin:2016axk}. In particular, the ``diagram removal with interference'' (DR2)
approach, as implemented in the {\sc MadSTR} plugin~\cite{Frixione:2019fxg}, is employed to subtract contributions from resonant top quarks which appear at NLO.
\end{itemize}

It is worth to notice that, since YR4, important theoretical progress has been achieved for these processes, in particular: soft-gluon resummation 
has been performed for $ttH$ at NNLL accuracy~\cite{Broggio:2015lya,Kulesza:2015vda,Broggio:2016lfj,Kulesza:2017ukk}, 
and it has been subsequently combined with the EW corrections~\cite{ Kulesza:2018tqz,Broggio:2019ewu,Kulesza:2020nfh}. More recently, the NNLO
cross section has been computed~\cite{Catani:2022mfv}, relying on
an approximation solely for the two-loop amplitude; finally, EW corrections
have been computed for $tH$ production~\cite{Pagani:2020mov}; their effect is rather small (3\%), and furthermore they cannot be computed
for the $s$-, $t$- and $tW$ channels separately.


\section*{$\boldsymbol{b\bar{b}H}$}
The $b\bar bH$ cross sections at different values of $m_H$ are reported in Table ~\ref{tab:bbH} and obtained with a linear interpolation of the YR4 cross sections available at $\sqrt{s}=13.5$ TeV and $\sqrt{s}=14 $ TeV (For $M_H=125.4$ GeV we used the numbers for 13 TeV.). These are obtained for both the YR4 values and for the NLO+NNLLpart+$y_by_t$ predictions that are based on the following results:

\begin{itemize}
    \item QCD corrections within the 4-flavour scheme (4FS) at NLO including the interference between top- and bottom-Yukawa induced contributions~\cite{Dittmaier:2003ej,Dawson:2003kb,Wiesemann:2014ioa}.
    \item QCD corrections within the 5-flavour scheme at NLO \cite{Balazs:1998sb,Dicus:1998hs} and NNLO~\cite{Harlander:2003ai}.
    \item Santander matching of the 4FS and 5FS \cite{Harlander:2011aa}.
    \item Fully matched results of the 4FS and 5FS using the FONLL \cite{Forte:2015hba,Forte:2016sja} and NLO+NNLLpart+$y_by_t$ \cite{Bonvini:2015pxa,Bonvini:2016fgf} methods.
\end{itemize}

Since YR4 there have been additional work on the determination of the N$^3$LO corrections to bottom-induced Higgs-boson production in the 5FS that has been matched to the 4FS consistently \cite{Duhr:2020kzd}. Another study pointed out the relevance or dominance of non bottom-Yukawa induced processes as $gg\to Hg^*\to Hb\bar b$ \cite{Deutschmann:2018avk} and several other competing processes that overwhelm the SM bottom-Yukawa induced contribution \cite{Pagani:2020rsg}. The disentanglement of the bottom-Yukawa induced contribution from the other competing processes, however, might be possible by machine-learning techniques \cite{Grojean:2020ech}. In general, the bottom-Yukawa induced cross sections are relevant for BSM scenarios with an enhanced bottom Yukawa coupling as e.g.~supersymmetric extensions or a 2HDM of type II.

\section*{Acknowledgments}
This work was done on behalf of the LHCHWG.
\newpage

\begin{table}[!h]
\begin{center}
\caption{ \label{tab:ggF} ggF cross sections at the LHC at 13.6 TeV and corresponding scale and PDF+$\alpha_\mathrm{s}$ uncertainties computed according to the PDF4LHC recommendation.}
\begin{tabular}{ |l|l|l|l|l|l|l|l|  }
 \hline
 \multicolumn{8}{|c|}{\bf ggF (N3LO QCD + NLO EW)} \\
 \hline
 $m_H$ [GeV]	& $\sigma$ [pb]  & +Th $\%$	& -Th $\%$	& TH Gaussian $\%$	& ±(PDF+$\alpha_{\mathrm{s}}) \%$ &	$\pm$PDF$\%$ & $\pm\alpha_{\mathrm{s}} \%$ \\
 \hline
120.00 & 5.611E+01 & +4.7 & -6.9 & ±4.0 & ±3.2 & ±1.9 & ±2.6\\ 
 120.50 & 5.571E+01 & +4.7 & -6.9 & ±4.0 & ±3.2 & ±1.9 & ±2.6\\ 
 121.00 & 5.531E+01 & +4.7 & -6.8 & ±3.9 & ±3.2 & ±1.9 & ±2.6\\ 
 121.50 & 5.490E+01 & +4.7 & -6.8 & ±3.9 & ±3.2 & ±1.9 & ±2.6\\ 
 122.00 & 5.451E+01 & +4.6 & -6.8 & ±3.9 & ±3.2 & ±1.9 & ±2.6\\ 
 122.50 & 5.412E+01 & +4.6 & -6.8 & ±3.9 & ±3.2 & ±1.9 & ±2.6\\ 
 123.00 & 5.374E+01 & +4.6 & -6.8 & ±3.9 & ±3.2 & ±1.9 & ±2.6\\ 
 123.50 & 5.336E+01 & +4.6 & -6.8 & ±3.9 & ±3.2 & ±1.9 & ±2.6\\ 
 124.00 & 5.298E+01 & +4.6 & -6.7 & ±3.9 & ±3.2 & ±1.9 & ±2.6\\ 
 124.10 & 5.290E+01 & +4.6 & -6.7 & ±3.9 & ±3.2 & ±1.9 & ±2.6\\ 
 124.20 & 5.283E+01 & +4.6 & -6.7 & ±3.9 & ±3.2 & ±1.9 & ±2.6\\ 
 124.30 & 5.275E+01 & +4.6 & -6.7 & ±3.9 & ±3.2 & ±1.9 & ±2.6\\ 
 124.40 & 5.268E+01 & +4.6 & -6.7 & ±3.9 & ±3.2 & ±1.9 & ±2.6\\ 
 124.50 & 5.260E+01 & +4.6 & -6.7 & ±3.9 & ±3.2 & ±1.9 & ±2.6\\ 
 124.60 & 5.253E+01 & +4.6 & -6.7 & ±3.9 & ±3.2 & ±1.9 & ±2.6\\ 
 124.70 & 5.245E+01 & +4.6 & -6.7 & ±3.9 & ±3.2 & ±1.9 & ±2.6\\ 
 124.80 & 5.238E+01 & +4.6 & -6.7 & ±3.9 & ±3.2 & ±1.9 & ±2.6\\ 
 124.90 & 5.231E+01 & +4.6 & -6.7 & ±3.9 & ±3.2 & ±1.9 & ±2.6\\ 
 125.00 & 5.223E+01 & +4.6 & -6.7 & ±3.9 & ±3.2 & ±1.9 & ±2.6\\ 
 125.09 & 5.217E+01 & +4.6 & -6.7 & ±3.9 & ±3.2 & ±1.9 & ±2.6\\ 
 125.10 & 5.216E+01 & +4.6 & -6.7 & ±3.9 & ±3.2 & ±1.9 & ±2.6\\ 
 125.20 & 5.209E+01 & +4.6 & -6.7 & ±3.9 & ±3.2 & ±1.9 & ±2.6\\ 
 125.30 & 5.202E+01 & +4.6 & -6.7 & ±3.9 & ±3.2 & ±1.9 & ±2.6\\ 
 125.38 & 5.196E+01 & +4.6 & -6.7 & ±3.9	& ±3.2 & ±1.9 & ±2.6\\
 125.40 & 5.194E+01 & +4.6 & -6.7 & ±3.9 & ±3.2 & ±1.9 & ±2.6\\ 
 125.50 & 5.187E+01 & +4.6 & -6.7 & ±3.9 & ±3.2 & ±1.9 & ±2.6\\ 
 125.60 & 5.180E+01 & +4.6 & -6.7 & ±3.9 & ±3.2 & ±1.9 & ±2.6\\ 
 125.70 & 5.172E+01 & +4.6 & -6.7 & ±3.9 & ±3.2 & ±1.9 & ±2.6\\ 
 125.80 & 5.165E+01 & +4.6 & -6.7 & ±3.9 & ±3.2 & ±1.9 & ±2.6\\ 
 125.90 & 5.158E+01 & +4.6 & -6.7 & ±3.9 & ±3.2 & ±1.9 & ±2.6\\ 
 126.00 & 5.151E+01 & +4.6 & -6.7 & ±3.9 & ±3.2 & ±1.9 & ±2.6\\ 
 126.50 & 5.115E+01 & +4.5 & -6.7 & ±3.8 & ±3.2 & ±1.9 & ±2.6\\ 
 127.00 & 5.080E+01 & +4.5 & -6.6 & ±3.8 & ±3.2 & ±1.9 & ±2.6\\ 
 127.50 & 5.045E+01 & +4.5 & -6.6 & ±3.8 & ±3.2 & ±1.9 & ±2.6\\ 
 128.00 & 5.011E+01 & +4.5 & -6.6 & ±3.8 & ±3.2 & ±1.9 & ±2.6\\ 
 128.50 & 4.976E+01 & +4.5 & -6.6 & ±3.8 & ±3.2 & ±1.9 & ±2.6\\ 
 129.00 & 4.943E+01 & +4.5 & -6.6 & ±3.8 & ±3.2 & ±1.9 & ±2.6\\ 
 129.50 & 4.909E+01 & +4.5 & -6.6 & ±3.8 & ±3.2 & ±1.9 & ±2.6\\ 
 130.00 & 4.875E+01 & +4.5 & -6.6 & ±3.8 & ±3.2 & ±1.8 & ±2.6\\
 \hline
\end{tabular}
\end{center}
\end{table}

\begin{table}[!h]
\begin{center}
\caption{ \label{tab:VBF} VBF cross sections at the LHC at 13.6 TeV and corresponding scale and PDF+$\alpha_\mathrm{s}$ uncertainties computed according to the PDF4LHC recommendation.}
\begin{tabular}{ |l|l|l|l|l|l|l|  }
 \hline
 \multicolumn{7}{|c|}{\bf VBF ( (approx.) NNLO QCD + NLO EW )} \\
 \hline
 $m_H$ [GeV]	& $\sigma$ [pb]  & +Scale  $\%$	& -Scale  $\%$	& ±(PDF+$\alpha_{\mathrm{s}}) \%$ &	$\pm$PDF$\%$ & $\pm\alpha_{\mathrm{s}} \%$ \\
 \hline
125	   &    4.078	&+0.5	&-0.3	&±2.1	&±2.1	&±0.5 \\
125.09 &	4.075	&+0.5	&-0.3	&±2.1	&±2.1	&±0.5 \\
125.38 &	4.067	&+0.5	&-0.3	&±2.1	&±2.1	&±0.5 \\
 \hline
\end{tabular}
\end{center}
\end{table}

\begin{table}[!h]
\begin{center}
\caption{ \label{tab:WH} WH cross sections at the LHC at 13.6 TeV and corresponding scale and PDF+$\alpha_\mathrm{s}$ uncertainties computed according to the PDF4LHC recommendation.}
\begin{tabular}{ |l|l|l|l|l|l|l|l|l|  }
 \hline
 \multicolumn{9}{|c|}{\bf $pp\to WH$ (NNLO QCD + NLO EW ) } \\
 \hline
 $m_H$ [GeV]	& $\sigma$ [pb]  & +Scale $\%$	& -Scale $\%$	& ±(PDF+$\alpha_{\mathrm{s}}) \%$ &	$\pm$PDF$\%$ & $\pm\alpha_{\mathrm{s}} \%$ & $W^+H$ [pb] & $W^-H$ [pb]\\
 \hline
125	   &    1.457	&+0.4	&-0.7	&±1.8	&±1.6	&±0.9	&0.8889 	&0.5677   \\
125.09 &	1.453	&+0.4	&-0.7	&±1.8	&±1.6	&±0.9	&0.8870  	&0.5664 \\
125.38 &	1.442	&+0.4	&-0.7	&±1.8	&±1.6	&±0.9	&0.8801	    &0.5620 \\
 \hline
\end{tabular}
\vspace{1cm}
\caption{ \label{tab:ZH} ZH cross sections at the LHC at 13.6 TeV and corresponding scale and PDF+$\alpha_\mathrm{s}$ uncertainties computed according to the PDF4LHC recommendation.}
\begin{tabular}{ |l|l|l|l|l|l|l|l| }
 \hline
 \multicolumn{8}{|c|}{\bf $pp\to ZH$ (NNLO QCD + NLO EW ) } \\
 \hline
 $m_H$ [GeV]	& $\sigma$ [pb]  & +Scale $\%$	& -Scale $\%$	& ±(PDF+$\alpha_{\mathrm{s}}) \%$ &	$\pm$PDF$\%$ & $\pm\alpha_{\mathrm{s}} \%$ & $\sigma(gg\to ZH)$ [pb]\\
 \hline
125	   &    9.439E-01	&+3.7	&-3.2	&±1.6	&±1.3	&±0.9	&1.360E-01   \\
125.09 &	9.422E-01	&+3.8	&-3.2	&±1.6	&±1.3	&±0.9	&1.359E-01 \\
125.38 &	9.361E-01	&+3.8	&-3.2	&±1.6	&±1.3	&±0.9	&1.347E-01 \\
 \hline
\end{tabular}
\end{center}
\end{table}

\begin{table}[!h]
\begin{center}
\caption{ \label{tab:ttH} $t\bar{t}H$ cross sections at the LHC at 13.6 TeV and corresponding scale and PDF+$\alpha_\mathrm{s}$ uncertainties computed according to the PDF4LHC recommendation.}
\begin{tabular}{ |l|l|l|l|l|l|l|  }
 \hline
 \multicolumn{7}{|c|}{\bf $\boldsymbol{t\bar{t}H}$ (NLO QCD + NLO EW) } \\
 \hline
 $m_H$ [GeV]	& $\sigma$ [pb]  & $+$Scale  $\%$	& $-$Scale  $\%$	& $\pm$(PDF+$\alpha_{\mathrm{s}}) \%$ &	$\pm$PDF$\%$ & $\pm\alpha_{\mathrm{s}} \%$ \\
 \hline
125	   &    $5.700 \cdot 10^{-1}$	& $+6.0$	
& $-9.3$	& $\pm 3.5$	& $\pm 3$	& $\pm 2$ \\
125.09 &	$5.688\cdot 10^{-1}$	& $+6.0$	
& $-9.3$	& $\pm 3.5$	& $\pm 3$	& $\pm 2$ \\
125.38 &	$5.638\cdot 10^{-1}$	& $+6.0$	
& $-9.3$	& $\pm 3.5$	& $\pm 3$	& $\pm 2$ \\
125.40 &	$5.634\cdot 10^{-1}$	& $+6.0$	
& $-9.3$	& $\pm 3.5$	& $\pm 3$	& $\pm 2$ \\
 \hline
\end{tabular}
\vspace{1cm}
\caption{ \label{tab:tH-t} $tH$ $t$-channel cross sections at the LHC at 13.6 TeV and corresponding scale and PDF+$\alpha_\mathrm{s}$ uncertainties computed according to the PDF4LHC recommendation. Sum of both $tH$ and $\bar{t}H$ cross sections is given.}
\begin{tabular}{ |l|l|l|l|l|l|l|  }
 \hline
 \multicolumn{7}{|c|}{\bf $\boldsymbol{tH}$ $\boldsymbol{t}$-chan (NLO QCD) } \\
 \hline
 $m_H$ [GeV]	& $\sigma$ [pb]  & $+$Scale  $\%$	& $-$Scale  $\%$	& $\pm$(PDF+$\alpha_{\mathrm{s}}) \%$ &	$\pm$PDF$\%$ & $\pm\alpha_{\mathrm{s}} \%$ \\
 \hline
125	   &    $8.362 \cdot 10^{-2}$	& $+6.5$	& $-14.8$	
& $\pm 3.7$	& $\pm 3.5$	& $\pm 1.2$ \\
125.09 &	$8.353 \cdot 10^{-2}$	& $+6.5$	& $-14.8$	
& $\pm 3.7$	& $\pm 3.5$	& $\pm  1.2$ \\
125.38 &	$8.320 \cdot 10^{-2}$	& $+6.5$	& $-14.8$	
& $\pm 3.7$	& $\pm 3.5$	& $\pm 1.2$ \\
125.40 &	$8.317 \cdot 10^{-2}$	& $+6.5$	& $-14.8$	
& $\pm 3.7$	& $\pm 3.5$	& $\pm 1.2$ \\
 \hline
\end{tabular}
\vspace{1cm}
\caption{ \label{tab:tH-s}  $tH$ $s$-channel cross sections at the LHC at 13.6 TeV and corresponding scale and PDF+$\alpha_\mathrm{s}$ uncertainties computed according to the PDF4LHC recommendation. Sum of both $tH$ and $\bar{t}H$ cross sections is given.}
\begin{tabular}{ |l|l|l|l|l|l|l|  }
 \hline
 \multicolumn{7}{|c|}{\bf $\boldsymbol{tH}$ $\boldsymbol{s}$-chan (NLO QCD)} \\
 \hline
 $m_H$ [GeV]	& $\sigma$ [pb]  & $+$Scale  $\%$	& $-$Scale  $\%$	& $\pm$(PDF+$\alpha_{\mathrm{s}}) \%$ &	$\pm$PDF$\%$ & $\pm\alpha_{\mathrm{s}} \%$ \\
 \hline
125	   &    $3.068 \cdot 10^{-3}$	
& $+2.4$	& $-1.7$	& $\pm 2.2$	& $\pm 2.2$	& $\pm 0.3$ \\
125.09 &	$3.064 \cdot 10^{-3}$	
& $+2.4$	& $-1.7$	& $\pm 2.2$	& $\pm 2.2$	& $\pm 0.3$ \\
125.38 &	$ 3.044 \cdot 10^{-3}$	
& $+2.4$	& $-1.7$	& $\pm 2.2$	& $\pm 2.2$	& $\pm 0.3$ \\
125.40 &	$3.040 \cdot 10^{-3}$	& $+2.4$	& $-1.7$	
& $\pm 2.2$	& $\pm 2.2$	& $\pm 0.3$ \\
 \hline
\end{tabular}
\vspace{1cm}
\caption{ \label{tab:tWH}  $tH$  $W$-associated cross sections at the LHC at 13.6 TeV and corresponding scale and PDF+$\alpha_\mathrm{s}$ uncertainties computed according to the PDF4LHC recommendation. Sum of both $tH$ and $\bar{t}H$ cross sections is given.}
\begin{tabular}{ |l|l|l|l|l|l|l|  }
 \hline
 \multicolumn{7}{|c|}{\bf $\boldsymbol{tWH}$ (NLO QCD)} \\
 \hline
 $m_H$ [GeV]	& $\sigma$ [pb]  & $+$Scale $\%$	& $-$Scale  $\%$	& $\pm$(PDF+$\alpha_{\mathrm{s}}) \%$ &	$\pm$PDF$\%$ & $\pm\alpha_{\mathrm{s}} \%$ \\
 \hline
125	       & $1.720 \cdot 10^{-2}$	& $+5.0$	& $-6.8$	
& $\pm 6.3$	& $\pm 6.1$	& $\pm 1.5$ \\
 \hline
\end{tabular}
\end{center}
\end{table}

\begin{table}[!h]
\begin{center}
\caption{ \label{tab:bbH} $b\bar bH$ cross sections at the LHC at 13.6 TeV and corresponding scale and PDF+$\alpha_\mathrm{s}$ uncertainties computed according to the PDF4LHC recommendation.}
\begin{tabular}{ |l|l|l|l|  }
 \hline
 \multicolumn{4}{|c|}{\bf $\boldsymbol{b\bar bH}$ (NNLO QCD in 5FS, NLO QCD in 4FS, Santander matching)} \\
 \hline
 $m_H$ [GeV]	& $\sigma$ [pb]  & +(Scale+PDF+$\alpha_{\mathrm{s}}$)$\%$	& $-$(Scale+PDF+$\alpha_{\mathrm{s}}$)$\%$ \\
 \hline
125.0  & 0.5269 & +20.1 & $-$24.0 \\
125.09 & 0.5257 & +20.3 & $-$24.0 \\
125.40 & 0.5213 & +20.4 & $-$24.1 \\
 \hline
\end{tabular}
\vspace*{0.5cm}

\begin{tabular}{ |l|l|l|l|l|  }
 \hline
 \multicolumn{5}{|c|}{\bf $\boldsymbol{b\bar bH}$ (NLO+NNLLpart+ybyt matching)} \\
 \hline
 $m_H$ [GeV]	& $\sigma$ [pb]  & ±(Ren.+Fact.+Matching Scales)$\%$ & +(PDF+$\alpha_s$+$m_b$)$\%$ & $-$(PDF+$\alpha_s$+$m_b$)$\%$ \\
 \hline
125	   & 0.568	& $\pm 5.0$ & +2.0 & $-$2.1 \\
125.09 & 0.566	& $\pm 5.1$ & +2.0 & $-$2.0 \\
125.40 & 0.563	& $\pm 4.9$ & +1.9 & $-$1.9 \\
\hline
\end{tabular}
\end{center}
\end{table}


\end{document}